\documentclass[pdflatex,sn-mathphys-num]{sn-jnl}

\usepackage{graphicx}%
\usepackage{multirow}%
\usepackage{amsmath,amssymb,amsfonts}%
\usepackage{amsthm}%
\usepackage{mathrsfs}%
\usepackage[title]{appendix}%
\usepackage{xcolor}%
\usepackage{textcomp}%
\usepackage{manyfoot}%
\usepackage{booktabs}%
\usepackage{algorithm}%
\usepackage{algorithmicx}%
\usepackage{algpseudocode}%
\usepackage{listings}%

\usepackage{tikz}
\usetikzlibrary{
  arrows.meta,
  positioning,
  fit,
  shapes.geometric,
  decorations.pathreplacing,
  patterns
} 
\usepackage{epstopdf} 
\usepackage{makecell} 

\usepackage{booktabs}
\usepackage{threeparttable} 
\usepackage{subcaption}

\usepackage{microtype}
\usepackage{xcolor}
\definecolor{linkblue}{RGB}{0,0,150}
 
\theoremstyle{thmstyleone}%
\theoremstyle{thmstyletwo}%

\theoremstyle{thmstylethree}%

\begin{document}

\title[APerformance-Driven Causal Signal Engineering for  Financial Markets under Non-Stationarity]{Performance-Driven Causal Signal Engineering for  Financial Markets under Non-Stationarity}


\author{\fnm{Lucas} \sur{A. Souza}}\email{lasouza@if.usp.br}



\affil{
\orgdiv{Independet Researcher}, 
\orgaddress{
\city{Divinópolis}, \postcode{35500-173}, \state{MG}, \country{Brazil}}}



\abstract{
We introduce a performance-driven framework for constructing \emph{strictly causal}
forward-oriented observables in strongly non-stationary time series.
The method combines a robustly normalized composite of heterogeneous indicators with
a causally computed derivative component, yielding a local phase-leading effect that
is amplified near regime transitions while remaining fully causal.
A hysteresis-based decision functional maps the observable into discrete system states,
with execution delayed by one step to preserve strict temporal ordering. 
Adaptation is achieved through a walk-forward scheme, in which model parameters are selected using rolling train--validation windows and subsequently applied out-of-sample. In this setting, the validation segment acts as an internal performance screen rather than as a statistical validation set, and no claims of generalization are inferred from it alone. 
The framework is evaluated on high-frequency financial time series as an experimentally accessible realization of a non-stationary complex system.
Under a controlled zero-cost setting, the resulting dynamics exhibit a pronounced
risk-reshaping effect, characterized by smoother trajectories and reduced drawdowns
relative to direct exposure, and should be interpreted as an upper bound on achievable
performance.
These results illustrate how causal signal engineering can generate anticipatory
structure in non-stationary systems without relying on non-causal information,
explicit horizon labeling, or high-capacity predictive models.
}




 
\maketitle 

\section{Introduction}

Financial markets are paradigmatic examples of non-stationary complex systems,
characterized by recurrent regime changes in volatility, trend persistence, and
liquidity. From the perspective of statistical physics, these regimes can be viewed
as metastable phases whose lifetimes and transitions carry economically relevant
information. Early empirical work formalized this view by modeling bull and bear
markets as persistent states with duration-dependent dynamics, showing that the
probability of regime termination depends on its age rather than being memoryless
\cite{lunde2004duration}.

While regime classification is valuable for ex-post analysis and risk assessment,
it does not resolve the ex-ante problem faced by trading systems: constructing
decision rules that operate online, using only information available at execution.
This difficulty is amplified at intraday horizons, where the data-generating process
is strongly non-stationary and market observables are endogenously coupled through
feedback, reflexivity, and microstructure effects. In such conditions, price dynamics
are often dominated by self-excitation and internal interactions instead of by
exogenous information flows, a behavior consistent with markets operating near a
critical regime \cite{hardiman2013critical}.

These features pose fundamental challenges to standard econometric approaches.
In non-stationary settings, classical linear models and Granger-causality-based
methods may produce unstable or misleading conclusions, as demonstrated in
\cite{papana2015detecting,papana2022identification}. More recent work has reframed
non-stationarity as a potential source of identifiability for causal discovery
\cite{sadeghi2025causal}. However, such methods are primarily designed for structural inference and hypothesis testing,
and are not tailored to the construction of operational, real-time decision functionals.

In parallel, machine-learning-based approaches have been increasingly applied to
high-frequency financial data. Deep architectures with explicit causal constraints,
such as dilated causal convolutions, have shown strong performance in volatility
forecasting \cite{moreno2024deepvol}, while reinforcement learning and deep hedging
frameworks aim to learn optimal trading or hedging policies directly from data
\cite{jaisson2022deep,buehler2019deep}. Despite their expressive power, these methods
typically rely on high-capacity models and large training sets, making them sensitive
to regime shifts and prone to overfitting in non-stationary environments
\cite{arian2024backtest}. Their interpretability and robustness under persistent
structural change therefore remain open questions.

An alternative line of research emphasizes explicit causality and regime awareness
in trading systems. Approaches based on optimal causal paths \cite{stubinger2019causalpaths}
or on regime-dependent portfolio allocation \cite{nystrup2017dynamic} highlight the
importance of respecting temporal ordering and structural heterogeneity. These methods,
however, generally rely on explicit regime identification or latent-state inference,
which introduces additional modeling assumptions and estimation error.

In this work, we take an operator-design perspective.
We construct a forward-oriented observable from past data only, and embed it in a walk-forward selection protocol so that execution is strictly out-of-sample.
Building on a recently proposed causal signal framework \cite{souza2025forward},
we consider a composite observable obtained from heterogeneous technical components,
robustly normalized and causally smoothed. While the original formulation demonstrated
that such observables can encode precursor information about regime transitions, its
performance was found to be strongly regime- and scale-dependent.

Here, we extend this construction by embedding the causal observable within a performance-driven, walk-forward selection framework. Model hyperparameters are selected exclusively from past data using rolling train--validation windows and then applied out-of-sample, preserving strict temporal ordering between information, signal construction, and execution. The validation block is interpreted as an internal performance control rather than as a statistical validation set, and is used to rank candidate configurations locally in time rather than to certify generalization. This procedure allows the decision functional to adapt to evolving market conditions without violating causality.

An empirical evaluation on high-frequency cryptocurrency markets illustrates how
optimization-aware causal observables can systematically reshape risk and return
characteristics, producing smoother equity dynamics and improved drawdown control.
More broadly, the framework provides a bridge between causal signal engineering and
performance-based adaptation in non-stationary complex systems.

The remainder of the paper is organized as follows.
Section~\ref{sec:methodology} introduces the causal signal construction and decision
functional. Section~\ref{sec:trend_objective} connects the forward-oriented observable
to horizon-based trend objectives. Section~\ref{sec:optimization} describes the
walk-forward optimization protocol. Section~\ref{sec:experiments} presents empirical
results, and Section~\ref{sec:conclusion} concludes.


\section{Methodology}\label{sec:methodology} 

The proposed framework defines a fully causal decision functional designed to operate
in strongly non-stationary environments with frequent regime changes.
We construct a forward-oriented observable by transforming heterogeneous technical indicators
into a single composite signal, whose parameters are selected directly through economic performance.

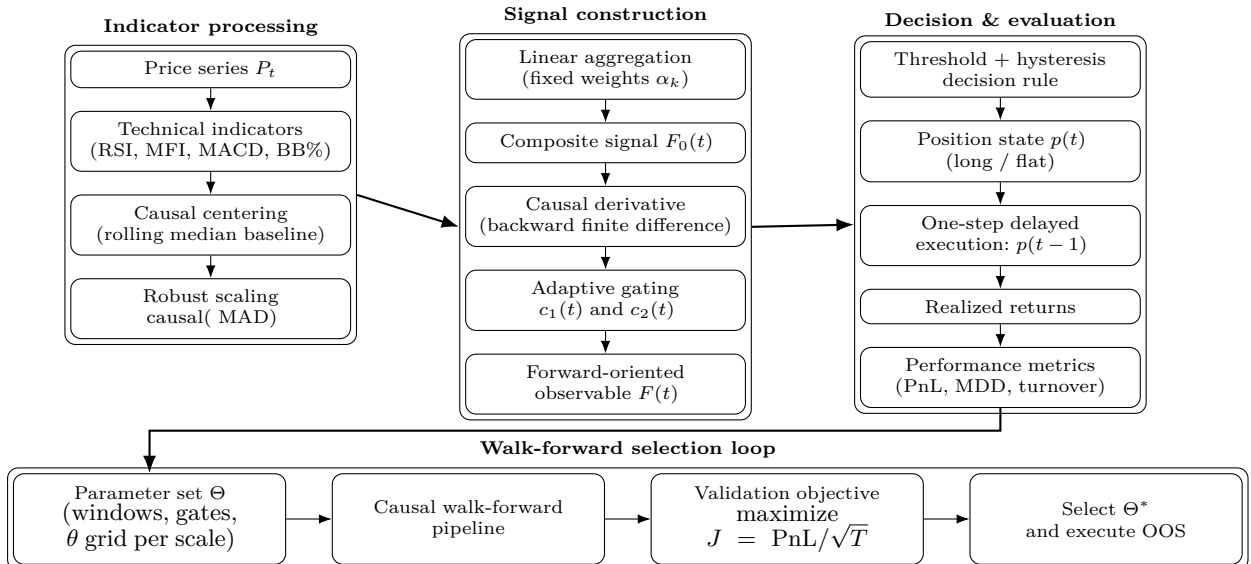
\begin{figure*}[!ht]
\centering

\begin{tikzpicture}[
node distance=3mm,
box/.style={rectangle, draw, rounded corners, align=center, inner sep=4pt, text width=3.4cm},
bigbox/.style={rectangle, draw, rounded corners, align=center, inner sep=2pt},
optblock/.style={rectangle, draw, rounded corners, align=center, inner sep=4pt, minimum width=3.3cm, minimum height=1.2cm, text width=3.3cm},
arrow/.style={-Latex},
font=\footnotesize
]

\node[box] (price) {Price series $P_t$};
\node[box, below=of price] (indicators) {Technical indicators\\(RSI, MFI, MACD, BB\%)};
\node[box, below=of indicators] (centering) {Causal centering\\(rolling median baseline)};
\node[box, below=of centering] (scale) {Robust scaling causal( MAD)};

\draw[arrow] (price) -- (indicators);
\draw[arrow] (indicators) -- (centering);
\draw[arrow] (centering) -- (scale);

\node[bigbox, fit=(price) (scale), label=above:{\bf Indicator processing}] (leftblock) {};

\node[box, right=15mm of price] (agg) {Linear aggregation\\(fixed weights $\alpha_k$)};
\node[box, below=of agg] (F0) {Composite signal $F_0(t)$};

\node[box, below=of F0] (deriv) {Causal derivative\\(backward finite difference)};
\node[box, below=of deriv] (gating) {Adaptive gating\\$c_1(t)$ and $c_2(t)$};
\node[box, below=of gating] (F) {Forward-oriented\\observable $F(t)$};

\draw[arrow] (agg) -- (F0);
\draw[arrow] (F0) -- (deriv); 
\draw[arrow] (deriv) -- (gating);
\draw[arrow] (gating) -- (F);

\node[bigbox, fit=(agg) (F), label=above:{\bf Signal construction}] (midblock) {};

\draw[arrow, thick] (leftblock.east) -- (midblock.west);

\node[box, right=15mm of agg] (thresh) {Threshold + hysteresis\\decision rule};
\node[box, below=of thresh] (state) {Position state $p(t)$\\(long / flat)};
\node[box, below=of state] (exec) {One-step delayed execution: $p(t\!-\!1)$};
\node[box, below=of exec] (returns) {Realized returns};
\node[box, below=of returns] (metrics) {Performance metrics\\(PnL, MDD, turnover)};

\draw[arrow] (thresh) -- (state);
\draw[arrow] (state) -- (exec);
\draw[arrow] (exec) -- (returns);
\draw[arrow] (returns) -- (metrics);

\node[bigbox, fit=(thresh) (metrics), label=above:{\bf Decision \& evaluation}] (rightblock) {};

\draw[arrow, thick] (midblock.east) -- (rightblock.west);

\node[optblock, below=7mm of midblock.south, xshift=-60mm] (paramset)
{Parameter set $\Theta$ \scriptsize (windows, gates, $\theta$ grid per scale)};

\node[optblock, right=6mm of paramset] (pipeline) {Causal walk-forward\\pipeline};

\node[optblock, right=6mm of pipeline] (objective)
{Validation objective\\\scriptsize maximize $J=\mathrm{PnL}/\sqrt{T}$ };

\node[optblock, right=6mm of objective] (update) {Select $\Theta^{*}$\\ and execute OOS};

\draw[arrow] (paramset) -- (pipeline);
\draw[arrow] (pipeline) -- (objective);
\draw[arrow] (objective) -- (update);

\draw[arrow, thick]
(metrics.south)
|- ++(0,-3mm)
-| (paramset.north);

\node[bigbox, fit=(paramset) (update),
label=above:{\bf Walk-forward selection loop}] (optbox) {};

\end{tikzpicture}
\caption{Pipeline overview. Prices are mapped to indicators, causally normalized, aggregated into $F_0(t)$, and transformed into a forward-oriented observable $F(t)$ via a gated derivative term. A hysteresis rule produces the binary exposure state, evaluated under a walk-forward protocol.}

\label{fig:work_diagram}
\end{figure*}

The methodology follows a modular pipeline (Fig.~\ref{fig:work_diagram}).
Raw prices are first mapped into standard indicators and processed through a robust
causal normalization that removes scale effects and mitigates outliers.
The normalized components are then aggregated and augmented by a derivative-based
term that introduces controlled anticipatory structure without violating causality.
The resulting signal is converted into trading actions via a hysteresis-based
decision rule and evaluated under a walk-forward, strictly out-of-sample protocol. 
All steps are causal; out-of-sample blocks are concatenated to form the global equity.

\subsection{Causal Robust Normalization}
\label{subsec:causal_norm}

All technical indicators are first mapped to a common dimensionless scale through a two-step causal transformation based on rolling medians and Median Absolute Deviation (MAD). Let \(w_{\text{norm}}\) denote the normalization window.

It is worth noting that the indicators selected for this work differ in scale: RSI, MFI, and BB$\%$ operate within a fixed range of [0, 100], facilitating a uniform interpretation. By contrast, the MACD has no predefined bounds and scale proportionally to both the asset price and its volatility. See Appendix~\ref{app:indicators} for details.
For each indicator \(I^{(k)}_t\) and for \(t \ge w_{\text{norm}}\), define the causal window
\[
\mathcal{W}_t(w_{\text{norm}}) = \{\tau : t - w_{\text{norm}} \le \tau < t\},
\]
containing the \(w_{\text{norm}}\) observations strictly prior to \(t\).
The causal rolling median (baseline) is
\[
m^{(k)}_t = \operatorname{median}\{ I^{(k)}_\tau : \tau \in \mathcal{W}_t(w_{\text{norm}}) \},
\]
and the centered series is
\[
\tilde I^{(k)}_t = I^{(k)}_t - m^{(k)}_t.
\]

A robust local scale is obtained via the causal MAD on the centered series:
\[
s^{(k)}_t = \operatorname{median}\{ |\tilde I^{(k)}_\tau| : \tau \in \mathcal{W}_t(w_{\text{norm}}) \} + \varepsilon,
\]
where \(\varepsilon > 0\) is a small constant preventing division by zero. Both \(m^{(k)}_t\) and \(s^{(k)}_t\) depend only on past data (\(\tau < t\)), ensuring strict causality.
The normalized indicator is then
\begin{equation}
\label{eq:normalized_indicator}
Z^{(k)}_t = \frac{I^{(k)}_t - m^{(k)}_t}{s^{(k)}_t}.
\end{equation}
This transformation yields a scale-free and outlier-robust representation of each
indicator.
All subsequent steps operate exclusively on the normalized series
$\{Z^{(k)}_t\}$, ensuring that differences in scale or volatility do not dominate
the composite signal.

\subsection{Composite Observable}

Given the normalized indicators $\{Z^{(k)}_t\}$, we define a composite observable
as a convex linear combination:
\begin{equation}
\label{eq:F0_def}
F_0(t) = \sum_{k=1}^{K} \alpha_k Z^{(k)}_t,
\end{equation}
where $\alpha_k \ge 0$ and $\sum_{k=1}^{K} \alpha_k = 1$.
This construction preserves the dimensionless nature of the normalized indicators
and yields a composite signal interpretable as a consensus Z-score.

In this study, the aggregation weights are fixed and chosen uniformly,
$\alpha_k = 1/K$.
This design choice is deliberate.
By holding the weights constant, we isolate the effect of the causal normalization, derivative-based enhancement, and regime gating mechanisms, reducing the risk of overfitting and improving interpretability. Allowing the weights to vary or be optimized is a natural extension of the framework and is left for future work.

We employ four market indicators\footnote{
Throughout this work, ``MACD'' refers to the MACD histogram (i.e., the difference between the MACD line and its signal line), and ``BB\%'' denotes the Bollinger \%B indicator, also commonly referred to as $\%b$ in the technical analysis literature.}:
$Z_t^{(\mathrm{RSI})}$,
$Z_t^{(\mathrm{MFI})}$,
$Z_t^{(\mathrm{BB}\%)}$, and
$Z_t^{(\mathrm{MACD})}$,
whose formal definitions are provided in Appendix~\ref{app:indicators}.
Each component captures a distinct aspect of market behavior:
RSI measures momentum extremes,
MFI incorporates volume-weighted pressure,
BB\% encodes volatility-adjusted price location,
and MACD reflects trend acceleration.
Their aggregation into $F_0(t)$ produces a multi-dimensional consensus signal.
Coherent deviations across indicators reinforce the composite magnitude, while
idiosyncratic noise from individual indicators is attenuated.
This structure provides a stable foundation for the forward-oriented transformation
introduced in the next subsection.

\subsection{Indicator Analysis}

The causal robust normalization introduced in
Section~\ref{subsec:causal_norm} maps each indicator into a dimensionless
representation that is stable under non-stationarity and suitable for linear
aggregation.

\begin{figure}[!ht]
\centering    
\includegraphics[width=0.75\linewidth]{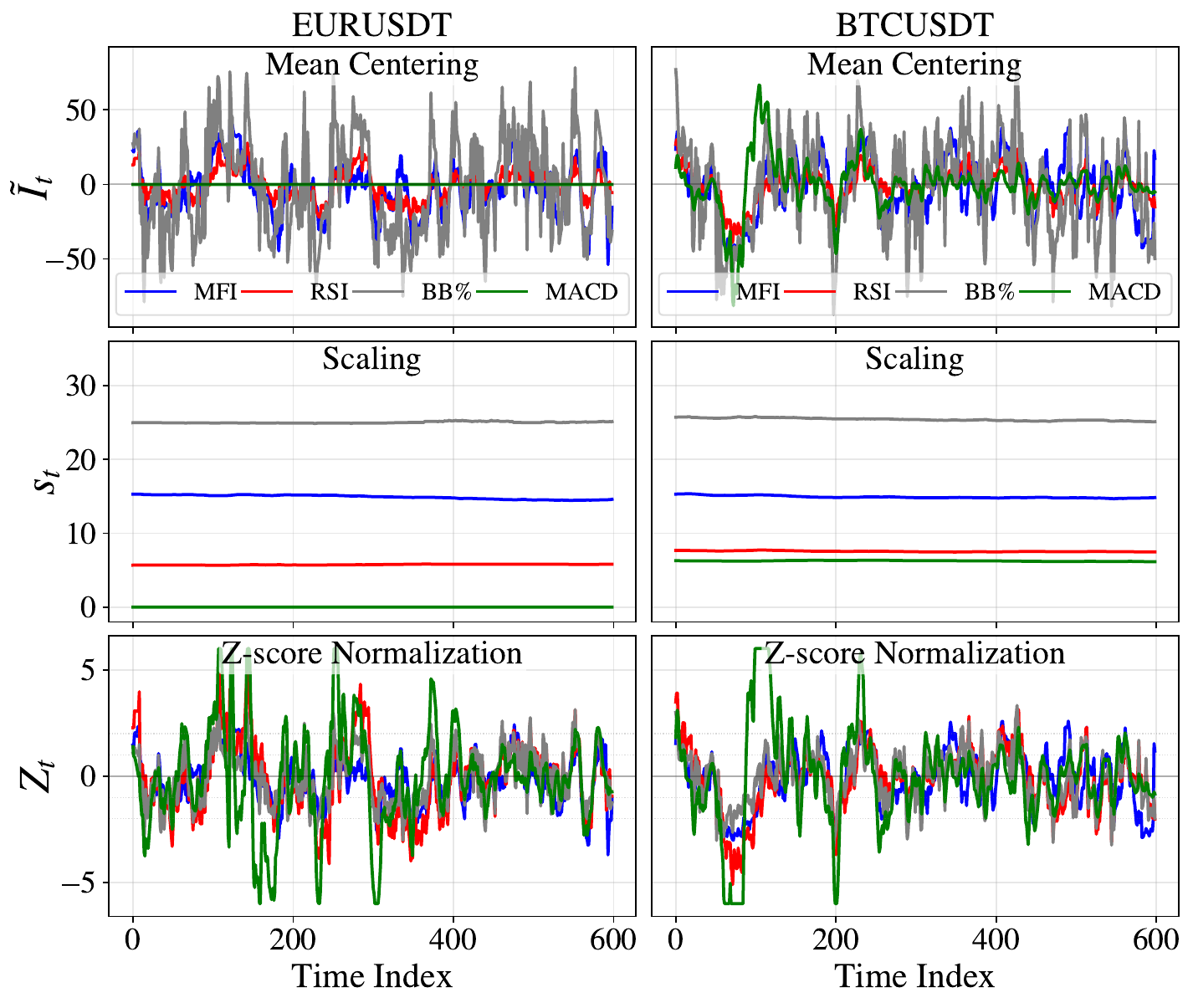}
\caption{
Causal robust normalization applied to four technical indicators (MFI, RSI, BB\%, and
MACD) for EURUSDT (left) and BTCUSDT (right).
\textbf{Top row:} centered indicators,
\( \tilde I^{(k)}_t = I^{(k)}_t - m^{(k)}_t \), where \( m^{(k)}_t \) denotes the causal
rolling median computed over a 5000-minute window.
\textbf{Middle row:} robust local scale estimates \( s^{(k)}_t \), obtained as the
rolling Median Absolute Deviation of the centered series.
\textbf{Bottom row:} normalized indicators,
\( Z^{(k)}_t = \tilde I^{(k)}_t / s^{(k)}_t \) (dimensionless), mapped to a common dimensionless scale,
enabling direct comparison and aggregation across indicators. }
\label{fig:indicators_analysis}
\end{figure}

Figure~\ref{fig:indicators_analysis} illustrates this transformation for the four
indicators used in this study (RSI, MFI, MACD, and BB\%) on EURUSDT and BTCUSDT,
using a 600-minute excerpt of one-minute data. The normalization window is set to
$w_{\text{norm}} = 5000$ minutes, balancing statistical robustness against
adaptation to evolving market conditions.
The top row shows the centered indicators, obtained by subtracting a causal rolling
median. This step removes slow-moving baselines and aligns all series around zero.
In raw units, however, the indicators operate on markedly different numerical
scales: RSI, MFI, and BB\% are bounded by construction, whereas the MACD is
scale-dependent, as it is defined as a difference of moving averages of the
underlying price.

For EURUSDT, prices are of order unity, and the associated moving averages therefore
remain close to one. Consequently, the MACD signal naturally lies at a much smaller
scale, of order $10^{-4}$, reflecting relative price variations rather than absolute
levels.
The middle row reports robust local scale estimates computed via the causal MAD. Since the MAD is not scale-invariant, the smaller values
observed for the MACD reflect its lower numerical scale rather than reduced
dynamical variability. In this example, the BB\% exhibits the largest dispersion,
followed by the MFI and the RSI.

After normalization (bottom row), all indicators fluctuate on comparable,
dimensionless scales. Expressed as deviations from a local median in units of
robust local variability, each component contributes symmetrically to the
composite observable $F_0(t)$, preventing dominance by any single indicator and
ensuring meaningful aggregation under non-stationarity.

\subsection{Forward-Looking Operator}
\label{sec:F_def}

To incorporate forward-oriented structure while preserving strict causality, we define
a derivative-enhanced observable \(F(t)\) from the composite signal \(F_0(t)\).
Let \(n_{\text{diff}}\) denote the lookback window used to compute causal finite
differences, and let \(w_{\text{MA}}\) be a short smoothing window applied to the
resulting difference series. The discrete \emph{causal} derivative is defined as
\[
\partial_t^{(n_{\text{diff}})} F_0(t)
= \frac{F_0(t) - F_0(t - n_{\text{diff}})}{n_{\text{diff}}},
\]
where \(n_{\text{diff}} \geq 1\) is the window length instead of the order of
differentiation.
Unlike the unit-lag differencing operator employed in ARIMA models
(see \citet{box1970time_Wilson2016} and \citet{hamilton1994time}), which is primarily designed to remove stochastic
trends and enforce stationarity, the present operator uses a finite backward difference
over an extended horizon to extract slope information that is informative about
near-future signal evolution. The resulting quantity approximates the local rate of
change of \(F_0(t)\) based exclusively on past observations.

To reduce high-frequency noise, the derivative series is further smoothed via a causal
moving average MA of length \(w_{\text{MA}}\),
\[
\widetilde{\partial_t F_0}(t)
= \mathrm{MA}_{w_{\text{MA}}}\!\left[ \partial_t^{(n_{\text{diff}})} F_0 \right]_t .
\]

The forward-oriented observable is then defined as
\begin{equation}
\label{eq:F_def}
F(t) = c_1(t)\,F_0(t) + c_2(t)\,\widetilde{\partial_t F_0}(t),
\end{equation}
with state-dependent mixing coefficients
\[
c_1(t) = \tanh\!\left(|\lambda_1 F_0(t)|\right);  \ 
c_2(t) = A\!\left[1 - \tanh\!\left(|\lambda_2 F_0(t)|\right)\right],
\]
where \(\lambda_1,\lambda_2 > 0\) control sensitivity to the magnitude of the composite
signal and \(A>0\) sets the overall scale of the derivative contribution.

In strongly persistent regimes, where \(|F_0(t)|\) is large, the coefficient \(c_2(t)\)
is suppressed and \(F(t)\) behaves similarly to the level signal \(F_0(t)\).
Conversely, near transition regions (\(F_0(t) \approx 0\)), the derivative term is
emphasized, yielding a causal approximation to a small forward shift of the composite
signal. The derivative-induced phase advance is illustrated in Appendix~\ref{app:derivative_leading}.
The operator does not forecast future values; it mixes level and slope information in real time. The link to horizon-based trend objectives is discussed in Section~\ref{sec:trend_objective}.


\subsection{Bridging to Horizon-Based Trend Objectives}
\label{sec:trend_objective}

The forward-oriented operator $F(t)$ is designed to causally anticipate changes in the
trajectory of the composite indicator $F_0(t)$. This objective is closely related to
the notion of trend persistence and regime duration that underpins a large body of
empirical work in financial economics. Seminal contributions by \citet{pagan2003simple}
and \citet{lunde2004duration} formalize market trends as bull and bear phases and study
their statistical properties ex post, including the duration dependence of regime
termination.

Our approach shifts the focus from ex-post classification to ex-ante anticipation.
We construct a causal signal that responds early to structural changes in market dynamics.
Empirically, the non-causal advance $F_0(t+6)$ exhibits strong alignment with a
forward-looking price-based trend metric, suggesting that the composite indicator
contains information relevant to future regime transitions.  

To formalize the trend target used in the literature, consider a price series
$\{P_t\}$ and define the horizon-based future and past price averages as
\[
\mu_H^{\mathrm{fut}}(t)
=
\frac{1}{H}\sum_{i=1}^{H} P_{t+i},
\qquad
\mu_H^{\mathrm{past}}(t)
=
\frac{1}{H}\sum_{i=0}^{H-1} P_{t-i}.
\]
The corresponding horizon-based percentage trend indicator is then given by
\begin{equation}
\Delta^{(H)}(t)
=
\frac{\mu_H^{\mathrm{fut}}(t)-\mu_H^{\mathrm{past}}(t)}{\mu_H^{\mathrm{past}}(t)}.
\label{eq:horizon_trend}
\end{equation}
which contrasts the average future price over the next $H$ periods with the average
past price over the preceding $H$ periods. This construction, or close variants of it,
is widely used as a benchmark for identifying local trend direction and strength.
 
\begin{figure}[!ht]
\centering
\includegraphics[width=0.6\linewidth]{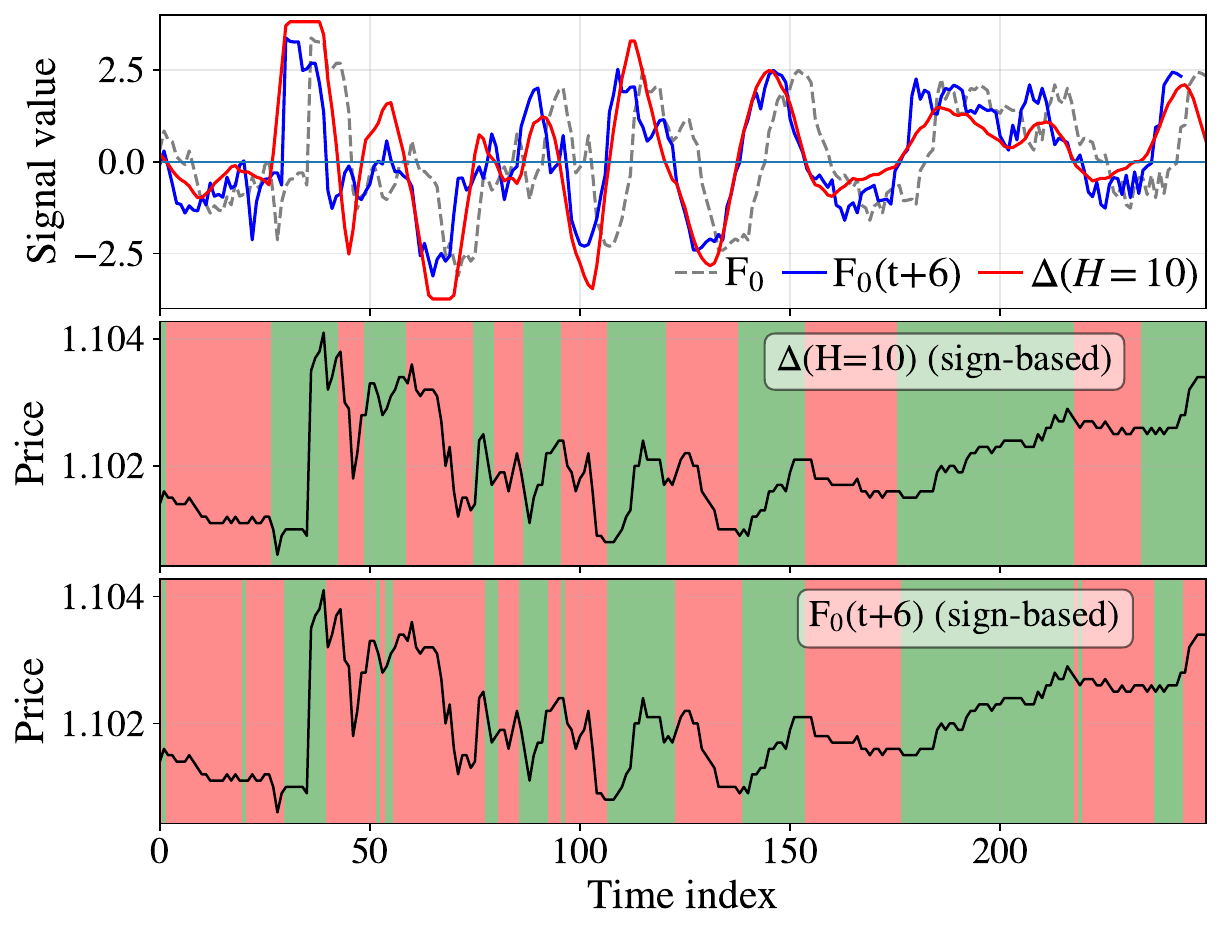}
\caption{Relationship between the composite indicator $F_0(t)$ (gray), its non-causal
advance $F_0(t+6)$ (blue), and the horizon-based trend indicator $\Delta^{(10)}(t)$
(red). The advanced signal exhibits alignment with the horizon trend, particularly
around zero-crossings associated with local regime transitions. Lower panels illustrate
the corresponding sign-based regime markings used for visualization.}
\label{fig:F0_vs_DeltaH}
\end{figure}

Figure~\ref{fig:F0_vs_DeltaH} illustrates the empirical relationship between the
contemporaneous composite signal $F_0(t)$, its non-causal advance $F_0(t+6)$, and the
horizon trend indicator $\Delta^{(10)}(t)$. Despite being derived from distinct
sources, $F_0(t+6)$ and $\Delta^{(10)}(t)$ exhibit similar zero-crossing behavior,
suggesting that advancing $F_0$ by a small number of steps captures information
relevant for horizon-based trend objectives. Naturally, $F_0(t+6)$ is not available
in real time and cannot be used operationally.

The two lower panels provide an illustrative mapping from these signals to directional
regimes, shown side by side on a shared time axis. For simplicity, buy (green) and sell
(red) regions are defined by sign changes of the respective signals, i.e., crossings
of a zero reference level. This visualization is intended solely to highlight the
similarity in regime timing induced by the two constructions and to serve as a
conceptual bridge to the explicit decision rules introduced in the next section.

The derivative component in the forward-oriented operator $F(t)$ provides a causal
mechanism to approximate this anticipatory behavior. 
This approximation implicitly assumes that the composite signal $F_0(t)$ admits a
locally smooth evolution, e.g., is locally Lipschitz-continuous over the relevant
time scale, so that a first-order expansion provides a meaningful description of
its short-horizon dynamics.
Since $\partial_t F_0(t)$ captures
the local rate of change of the composite signal, a first-order expansion yields
\begin{equation}
F_0(t+\delta) \approx F_0(t) + \delta\,\partial_t F_0(t),
\end{equation}
showing that a suitably scaled derivative term can emulate a small forward shift.
In our construction, the adaptive gating coefficient $c_2(t)$ effectively implements a
regime-dependent $\delta$, amplifying the derivative contribution near transitions and
suppressing it during strong trending phases.

Although a fixed multi-step advance (e.g., six periods) is impossible under strict
causality, the derivative-enhanced signal $F(t)$ acts as a causal surrogate for limited
forward anticipation. This provides a direct conceptual link between the proposed
signal construction and horizon-based trend targets widely used in the literature.
The economic relevance of this surrogate is ultimately assessed through the
PnL-driven walk-forward optimization described in
Section~\ref{sec:optimization}.
  


\subsection{Decision Rule and Performance Measurement}
\label{subsec:decision_rule}

The forward-oriented observable \(F(t)\) serves as the primary input to a binary decision rule. Let \(s_t = F(t)\) denote the signal value at time \(t\). A position state \(p_t \in \{0,1\}\) is maintained, where \(p_t = 1\) corresponds to a long position and \(p_t = 0\) to a flat (cash) position.

To mitigate noise-induced whipsaws, state transitions are governed by a hysteresis mechanism with threshold \(\theta>0\):
\begin{equation}
p_t =
\begin{cases}
1, & \text{if } p_{t-1} = 0 \text{ and } s_t > +\theta, \\
0, & \text{if } p_{t-1} = 1 \text{ and } s_t < -\theta, \\
p_{t-1}, & \text{otherwise}.
\end{cases}
\label{eq:hysteresis_rule}
\end{equation}

This design ensures that a switch from flat to long occurs only when the signal surpasses a positive threshold, while a switch from long to flat requires the signal to drop below a negative threshold. The neutral zone between \(-\theta\) and \(+\theta\) prevents excessive trading when the signal oscillates near zero.

For a strictly causal implementation, the position state used to compute returns at time \(t\) is the previous state \(p_{t-1}\). Let \(r_t = (P_t - P_{t-1})/P_{t-1}\) be the simple return of the underlying asset. The strategy return at time \(t\) is then:
\begin{equation}
R_t = p_{t-1} \cdot r_t.
\label{eq:strategy_return}
\end{equation}

The cumulative equity curve is obtained by compounding:
\begin{equation}
V_t = V_0 \prod_{\tau=1}^{t} (1 + R_\tau),
\end{equation}
with \(V_0 = 1\) as the initial capital. Trading activity is quantified by the total number of position changes over a given horizon \(T\):
\begin{equation}
\text{Turnover}(T) = \sum_{t=1}^{T} |p_t - p_{t-1}|.
\end{equation}

\subsection{Signal Scaling and Gating Effects}
\label{subsec:signal_scaling_gating}
Before introducing the parameter optimization framework, we examine how the scale of
the constructed signal $F(t)$ depends on the gating parameters. This analysis is
essential because trading decisions are based on thresholding $F(t)$: the frequency of
trades and effective exposure depend on how often $|F(t)|$ exceeds a decision threshold
$\theta$. If the amplitude of $F(t)$ varies substantially with
$(\lambda_1,\lambda_2,A)$, jointly optimizing these parameters together with $\theta$
becomes ill-posed, as distinct configurations may generate similar trading behavior
through simple rescaling. Characterizing the dependence of the signal magnitude on the
gating parameters is therefore necessary to decouple signal construction from decision
calibration.

To isolate gating effects, we fix equal indicator weights $\alpha_k = 1/K$ and compute
$F(t)$ over the last 100{,}000 observations (approximately 70 days of one-minute data)
for EURUSDT and BTCUSDT. 
Price data consist of one-minute OHLCV series obtained via the Binance public Spot API
\cite{binance_api}.
The sample spans January 2022 to December 2025.
Missing candles are handled by forward-filling prices and setting volume to zero,
ensuring a regular time grid without introducing look-ahead bias.

The derivative parameters are fixed at
$n_{\text{diff}}=2$ and $w_{\text{ma}}=2$. For each experiment, one gating parameter is
varied while the others are held constant, and the signal scale is summarized by
$\mathrm{median}(|F(t)|)$.

\begin{figure}[!ht]
\centering
\includegraphics[width=0.8\linewidth]{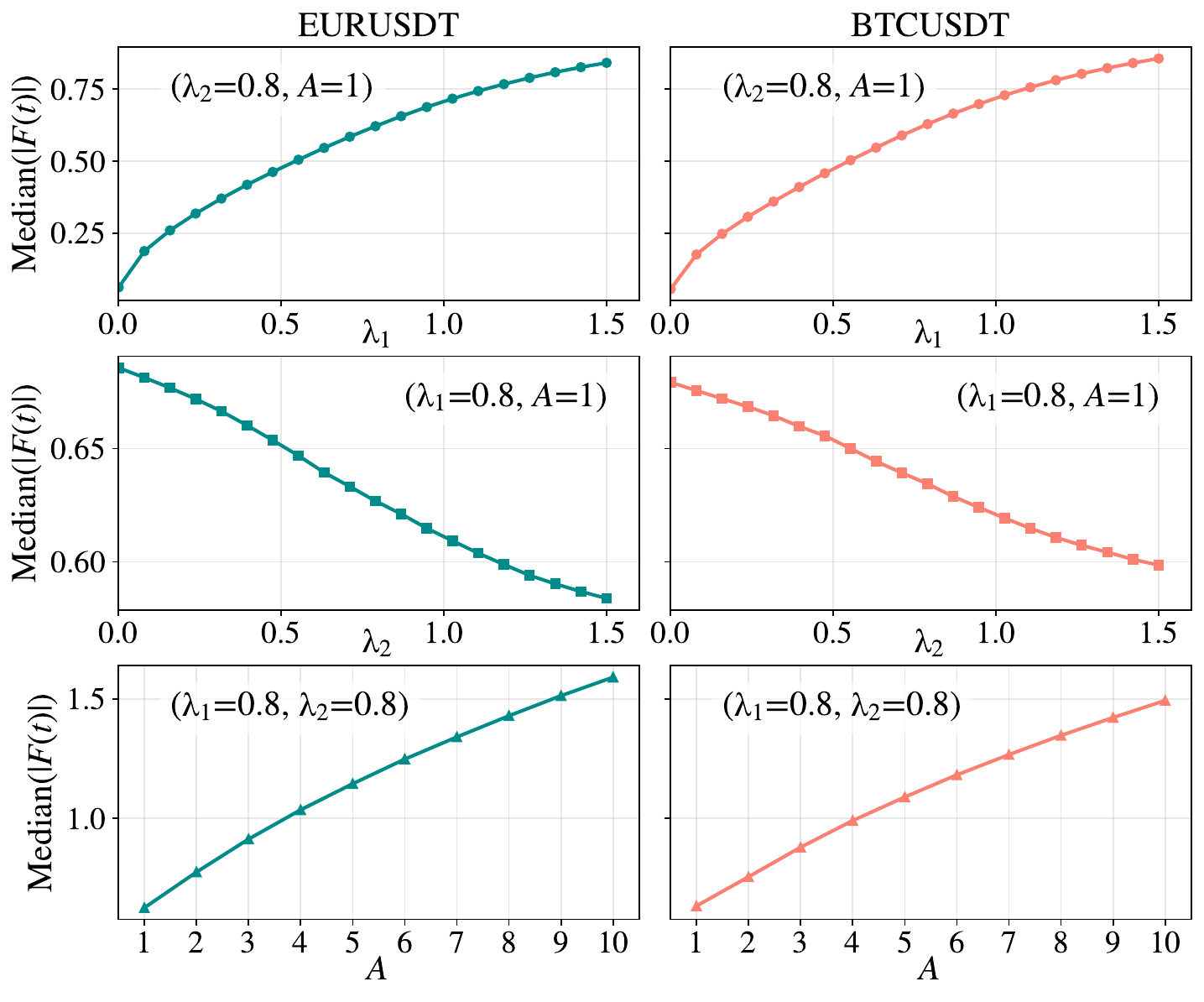}
\caption{Median magnitude of the constructed signal, $\mathrm{median}(|F(t)|)$, as a
function of the gating parameters $(\lambda_1,\lambda_2,A)$, computed over the last
100{,}000 observations for EURUSDT and BTCUSDT. The derivative parameters are fixed at
$n_{\text{diff}}=2$ and $w_{\text{ma}}=2$. The smooth and monotonic dependence indicates
that the gating parameters predominantly determine the scale of $F(t)$.}
\label{fig:median_F_vs_parameters}
\end{figure}

Figure~\ref{fig:median_F_vs_parameters} shows a smooth and monotonic dependence of the
signal magnitude on the gating parameters. Increasing $\lambda_1$ raises
$\mathrm{median}(|F(t)|)$ by strengthening the level component, while increasing
$\lambda_2$ suppresses the derivative contribution and reduces the overall amplitude.
The parameter $A$ scales the derivative channel almost linearly. These patterns are
consistent across both assets, indicating that the robust normalization of $F_0(t)$
stabilizes the baseline scale, whereas $(\lambda_1,\lambda_2,A)$ primarily control the
effective amplitude of $F(t)$.

Additional experiments (not shown) indicate that moderate variations around
$n_{\text{diff}}=2$ and $w_{\text{ma}}=2$ do not materially affect the structure or scale
of $F(t)$ relative to the dominant influence of the gating parameters. This separation
motivates treating the decision threshold $\theta$ independently of signal-shaping
parameters. The next section introduces the walk-forward optimization framework and
describes how $\theta$ is selected in a manner consistent with the observed scaling
behavior of $F(t)$.

\subsection{Parameter Optimization Framework}
\label{sec:optimization}
Given the forward-oriented signal $F(t)$ and the decision rule defined in
Section~\ref{subsec:decision_rule}, the trading strategy depends on a finite set of
parameters that govern both signal construction and temporal adaptation.
For clarity, we partition these parameters into two groups.

\begin{itemize}
    \item \textbf{Signal construction parameters}, which determine the shape and scale
    of the forward-oriented observable $F(t)$.
    \item \textbf{Walk-forward window parameters}, which control how the model adapts
    to non-stationarity over time.
\end{itemize}

The signal construction parameters are collected in
\begin{equation}
\Theta_{\text{sig}} = \bigl\{ n_{\text{diff}},\, w_{\text{ma}},\, \lambda_1,\, \lambda_2,\, A \bigr\},
\label{eq:signal_parameters}
\end{equation}
where $n_{\text{diff}}$ and $w_{\text{ma}}$ define the causal derivative operator, and
$(\lambda_1,\lambda_2,A)$ govern the adaptive gating mechanism introduced in
Eq.~\eqref{eq:F_def}. Threshold selection is treated separately (Section~\ref{subsec:theta_selection}) to avoid confounding signal scaling with execution filtering.
The walk-forward protocol introduces two additional window parameters,
\begin{equation}
\Theta_{\text{win}} = \bigl\{ w_{\text{fit}},\, \rho \bigr\},
\label{eq:window_parameters}
\end{equation}
where $w_{\text{fit}}$ denotes the length of the training window and $\rho$ determines
the validation horizon via $w_{\text{val}} = \lfloor w_{\text{fit}}/\rho \rfloor$.
The out-of-sample execution window is set equal to the validation window
($w_{\text{exec}} = w_{\text{val}}$), ensuring a consistent evaluation horizon across
epochs, as detailed in Section~\ref{subsec:wf_protocol}.

At each walk-forward epoch, the parameter set optimized is therefore
\[
\Theta = \Theta_{\text{sig}} \cup \Theta_{\text{win}}.
\]
The optimization is performed via a discrete grid search over predefined candidate
values. The specific grids employed in the empirical analysis are reported in
Table~\ref{tab:param_grid}.

\paragraph{Objective function.}
For a given parameter configuration $\Theta$, the pipeline generates a sequence of
strategy returns $\{R_t(\Theta)\}_{t=1}^{T_{\text{eval}}}$.
The terminal profit-and-loss over the evaluation window is
\begin{equation}
\text{PnL}(\Theta) = \prod_{t=1}^{T_{\text{eval}}} \bigl(1 + R_t(\Theta)\bigr) - 1,
\end{equation}
assuming unit initial capital.
Here, $T_{\text{eval}}$ denotes the number of observations in the evaluation block on
which the objective is computed. 
To enable comparisons across evaluation windows of different lengths, we maximize the
time-normalized objective
\begin{equation}
J(\Theta) = \frac{\text{PnL}(\Theta)}{\sqrt{T_{\text{eval}}}},
\label{eq:objective_sqrtT}
\end{equation}
which penalizes configurations that achieve high PnL over very short horizons. This objective should be interpreted as a volatility-penalized growth proxy, not a risk-adjusted estimator. 
When multiple parameter sets attain similar objective values, turnover—measured as
the total number of position changes over the evaluation window—is used as a
tie-breaking criterion to favor more stable trading behavior.

It is important to emphasize that this optimization should be interpreted as an internal, locally defined performance screen rather than as a mechanism for establishing statistically validated generalization. Because a finite grid of configurations is evaluated on each validation block, some degree of performance overfitting to local conditions is unavoidable. The role of the optimization is therefore not to identify a globally optimal or universally generalizing parameter set, but to maintain an internally consistent ranking of a small family of candidate configurations as market conditions evolve.

\subsection{Walk-Forward Protocol}
\label{subsec:wf_protocol}

Because the objective $J(\Theta)$ is evaluated through a fully causal backtest, gradient information is unavailable and parameter selection is performed via black-box optimization. To respect non-stationarity while preserving strict causality, we adopt a walk-forward protocol in which \emph{estimation, model selection, and execution are explicitly separated in time}. In this protocol, the “validation” block serves purely as an operational performance filter that ranks candidate configurations based on recent behavior; it is not interpreted as an independent hold-out set providing formal guarantees about future generalization.

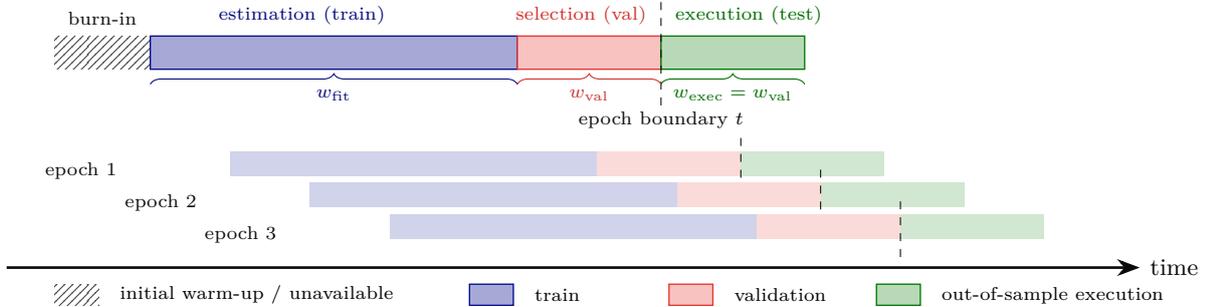
\begin{figure*}[!ht]
\centering
\begin{tikzpicture}[x=1.05cm,y=1.0cm, font=\small]

  \definecolor{cEst}{RGB}{0,10,139}   
  \definecolor{cVal}{RGB}{229,52,43}   
  \definecolor{cExe}{RGB}{0,120,0}  
  \definecolor{cBurn}{RGB}{70,70,70} 
  \definecolor{cAxis}{RGB}{10,10,10}

  \def\B{1.2}    
  \def\TR{4.6}   
  \def\VA{1.8}   
  \def\TE{1.8}   
  \def\SHIFT{1.0} 

  \draw[shift={(0,-0.45)}][-{Stealth[length=3mm]}, line width=1, color=cAxis] (0,0) -- (14.2,0)
    node[anchor=west, color=cAxis] {time}; 

  \def\x0{0.6}
  \def\y0{2.4}

  \fill[cBurn, pattern=north east lines, pattern color=cBurn]
    (\x0,\y0-0.22) rectangle ++(\B,0.44);
  \node[anchor=south, color=cAxis] at (\x0+0.6,\y0+0.25) {\footnotesize burn-in};

  \fill[cEst!35] (\x0+\B,\y0-0.22) rectangle ++(\TR,0.44);
  \draw[color=cEst, line width=0.5pt] (\x0+\B,\y0-0.22) rectangle ++(\TR,0.44);
  \node[anchor=south, color=cEst!90!black] at (\x0+1.8+\B+0.1,\y0+0.25) {\footnotesize estimation (train)};

  \fill[cVal!30] (\x0+\B+\TR,\y0-0.22) rectangle ++(\VA,0.44);
  \draw[color=cVal, line width=0.5pt] (\x0+\B+\TR,\y0-0.22) rectangle ++(\VA,0.44);
  \node[anchor=south, color=cVal!90!black] at (\x0+0.7+\B+\TR+0.1,\y0+0.25) {\footnotesize selection (val)};

  \fill[cExe!28] (\x0+\B+\TR+\VA,\y0-0.22) rectangle ++(\TE,0.44);
  \draw[color=cExe, line width=0.5pt] (\x0+\B+\TR+\VA,\y0-0.22) rectangle ++(\TE,0.44);
  \node[anchor=south, color=cExe!90!black] at (\x0+1.+\B+\TR+\VA+0.1,\y0+0.25) {\footnotesize execution (test)};

  \draw[dashed, color=cAxis] (\x0+\B+\TR+\VA, \y0-0.7) -- (\x0+\B+\TR+\VA, \y0+0.75);
  \node[anchor=north, color=cAxis] at (\x0+\B+\TR+\VA, \y0-0.65) {\footnotesize epoch boundary \(t\)};

  \draw[decorate, decoration={brace, amplitude=4pt}, color=cEst!90!black]
    (\x0+\B,\y0-0.42) -- (\x0+\B+\TR,\y0-0.42)
    node[midway, yshift=-4pt, color=cEst!90!black] {\footnotesize \(w_{\text{fit}}\)};
  \draw[decorate, decoration={brace, amplitude=4pt}, color=cVal!90!black]
    (\x0+\B+\TR,\y0-0.42) -- (\x0+\B+\TR+\VA,\y0-0.42)
    node[midway, yshift=-4pt, color=cVal!90!black] {\footnotesize \(w_{\text{val}}\)};
  \draw[decorate, decoration={brace, amplitude=4pt}, color=cExe!90!black]
    (\x0+\B+\TR+\VA,\y0-0.42) -- (\x0+\B+\TR+\VA+\TE,\y0-0.42)
    node[midway, yshift=-4pt, color=cExe!90!black] {\footnotesize \(w_{\text{exec}}=w_{\text{val}}\)};

  \foreach \k in {1,2,3} {
    \pgfmathsetmacro{\xx}{\x0 + \k*\SHIFT}
    \pgfmathsetmacro{\yy}{1.35 - 0.42*\k}

    \fill[cEst!18] (\xx+\B,\yy-0.16) rectangle ++(\TR,0.32);
    \fill[cVal!18] (\xx+\B+\TR,\yy-0.16) rectangle ++(\VA,0.32);
    \fill[cExe!18] (\xx+\B+\TR+\VA,\yy-0.16) rectangle ++(\TE,0.32);

    \draw[dashed, color=cAxis] (\xx+\B+\TR+\VA,\yy-0.4) -- (\xx+\B+\TR+\VA,\yy+0.4);

    \node[anchor=east, color=cAxis] at (\xx-0.1,\yy-0.1) {\footnotesize epoch \(\k \)};
  }

  \begin{scope}[shift={(0.6,-0.95)}]
    \fill[cBurn, pattern=north east lines, pattern color=cBurn] (0,0) rectangle (0.55,0.28);
    \node[anchor=west] at (0.7,0.14) {\footnotesize initial warm-up / unavailable};

    \fill[cEst!35] (5.2,0) rectangle (5.75,0.28);
    \draw[color=cEst, line width=0.5pt] (5.2,0) rectangle (5.75,0.28);
    \node[anchor=west] at (5.9,0.14) {\footnotesize train};

    \fill[cVal!30] (7.7,0) rectangle (8.25,0.28);
    \draw[color=cVal, line width=0.5pt] (7.7,0) rectangle (8.25,0.28);
    \node[anchor=west] at (8.4,0.14) {\footnotesize validation};

    \fill[cExe!28] (10.3,0) rectangle (10.85,0.28);
    \draw[color=cExe, line width=0.5pt] (10.3,0) rectangle (10.85,0.28);
    \node[anchor=west] at (11.0,0.14) {\footnotesize out-of-sample execution};
  \end{scope}

\end{tikzpicture}
\caption{Rolling window schedule used for causal walk-forward model selection. At each epoch boundary \(t\),
parameters are estimated on a past training block, selected using the immediately preceding validation block,
and then executed out-of-sample on the next block. We impose \(w_{\text{exec}}=w_{\text{val}}\) so that the
validation horizon matches the operational trading horizon, enabling directly comparable performance metrics.}
\label{fig:wf_windows}
\end{figure*}

At each epoch, the data are partitioned into three contiguous past-to-present
blocks (Figure~\ref{fig:wf_windows}):

\begin{itemize}
\item a \emph{training} block of length \(w_{\text{fit}}\), used exclusively for
signal construction and state estimation;
\item a \emph{validation} block of length \(w_{\text{val}}\), used solely for
internal model selection and short-horizon monitoring; it is not used to report
performance and does not constitute an out-of-sample test;
\item a subsequent \emph{test/forecast} block of length \(w_{\text{exec}}\), on which
the selected strategy is executed out-of-sample.
\end{itemize}

At an epoch boundary \(t\), these blocks are defined as
\begin{align}
\text{train:}\;& [\,t - w_{\text{fit}} - w_{\text{val}},\; t - w_{\text{val}}\,), \\
\text{val:}\;&   [\,t - w_{\text{val}},\; t\,), \\
\text{test:}\;&  [\,t,\; t + w_{\text{exec}}\,).
\end{align}

Crucially, parameter selection never uses information from the test block.
The parameter vector \(\Theta\) is chosen by maximizing the validation score
computed exclusively on the validation block:
\begin{equation}
\Theta^{*}(t) \;=\; \arg\max_{\Theta \in \mathcal{G}} 
J_{\text{val}}(\Theta;\, w_{\text{val}}),
\end{equation}
where \(\mathcal{G}\) denotes a finite grid of candidate configurations
(Table~\ref{tab:param_grid}).
The selected configuration \(\Theta^{*}(t)\) is then frozen and applied
\emph{without modification} to the subsequent test block, producing the
out-of-sample position sequence used to construct the global equity curve.
A schematic overview of the rolling train--validation--execution schedule,
including the enforced horizon constraint \(w_{\text{exec}} = w_{\text{val}}\),
is shown in Figure~\ref{fig:wf_windows}.

In addition to optimizing \(\Theta\), the protocol allows the window lengths
\((w_{\text{fit}}, w_{\text{val}}, w_{\text{exec}})\) themselves to vary across epochs.
Operationally, this is implemented as a discrete search over \(w_{\text{fit}}\) and
a ratio parameter \(\rho\), which determines the validation length via
\begin{equation}
w_{\text{val}} = \mathrm{round}\!\left(\frac{w_{\text{fit}}}{\rho}\right).
\label{eq:wval_from_ratio}
\end{equation}

We impose the constraint
\begin{equation}
w_{\text{exec}} = w_{\text{val}},
\label{eq:wpred_equals_wval}
\end{equation}
which ensures that model selection and out-of-sample execution are evaluated over
identical horizons.
This constraint serves three purposes. First, it enforces a clear separation
between estimation and execution while maintaining a consistent evaluation scale.
Second, it reduces the effective dimensionality of the window search, mitigating
the risk of horizon overfitting. Third, it yields a transparent operational
interpretation in which the trading horizon is calibrated on the most recent
validation period and immediately deployed on the next unseen block of the same
duration. Under this walk-forward scheme, the loop advances in steps of
$w_{\text{exec}}$, producing non-overlapping out-of-sample test blocks whose
concatenated positions define the global equity.

At this point, it is useful to contrast the present framework with earlier heuristic
constructions in the literature. That study can be recovered as a particular and more restrictive instance of the current construction, with simplified preprocessing and empirically chosen, fixed hyperparameters.
Related approaches based on derivative-augmented indicators have been explored in prior
work (see, e.g.,~\cite{souza2025forward}), but without the robust normalization and
systematic walk-forward calibration introduced here.  Aggregation weights were
adjusted to yield signals of similar visual amplitude, and key hyperparameters--
including the derivative gain, smoothing horizon, and mixing coefficients--were
selected empirically based on qualitative inspection and preliminary performance.

While this heuristic approach proved effective and provided early evidence of the
benefits of derivative-augmented signals, it lacked an explicit normalization
scheme, a principled calibration procedure, and a unified parameterization across
indicators. The present work addresses these limitations by introducing robust
MAD-based normalization, a formally defined composite signal scale, and a
systematic walk-forward calibration of all hyperparameters. Under this
formulation, the framework of \cite{souza2025forward} emerges as a special case
obtained by fixing the normalization to simple centering and selecting constant
parameter values, thereby situating the earlier results within a broader and more
robust causal signal engineering paradigm.


\section{Experimental Results}
\label{sec:experiments}

At intraday horizons, the design and evaluation of trading strategies face
substantial challenges due to the dominance of noise and quickly evolving
structural changes, a point long emphasized in the empirical market
microstructure literature, notably by~\citet{christensen2014fact}.
Cryptocurrency markets amplify these difficulties, exhibiting pronounced
non-stationarity, frequent regime shifts, and distinctive microstructural
dynamics, as documented by recent studies such as
\citet{easley2024microstructure} and \citet{deblasis2022arbitrage}.

Consistent with recent algorithmic trading literature that adopts rolling and
walk-forward evaluation protocols on high-frequency cryptocurrency data
(see, e.g.,~\citet{stefaniuk2025informer} and \citet{arian2024backtest}),
we evaluate the proposed framework on two liquid spot cryptocurrency markets,
EURUSDT and BTCUSDT, using one-minute sampled data.
Specifically, we assess a fully causal decision functional optimized in a
walk-forward fashion and evaluated at the one-minute frequency.

\subsection{Experimental Setup and Walk-Forward Protocol}
\label{subsec:experimental_setup}
 
All stages of the pipeline—indicator processing, normalization, signal
construction, parameter selection, and trading execution—are strictly causal,
relying exclusively on information available up to time $t$.
To isolate the intrinsic behavior of the signal construction and decision
mechanism, all results reported in this section are obtained under
\emph{zero transaction costs}. This choice allows us to analyze the raw
statistical and dynamical properties of the strategy independently of
market-specific frictions, which are addressed in a separate robustness
analysis.

Under the walk-forward framework introduced earlier, at each rebalancing time
$t$, model parameters are estimated through a train--validation split relying
exclusively on historical information (Section~\ref{sec:optimization}), after
which they are held fixed and applied out-of-sample over a contiguous test
window.
The global equity curve is constructed by concatenating the corresponding
out-of-sample position paths across successive windows.
The walk-forward optimization framework described in Section~\ref{sec:optimization} was implemented with the following parameter grids:

\begin{table}[!ht]
\centering
\footnotesize
\caption{Parameter grids used in the walk-forward selection.}
\label{tab:param_grid}
\renewcommand{\arraystretch}{0.95}
\setlength{\tabcolsep}{1pt}
\begin{tabular}{l l l}
\toprule
Param. & Description & Grid values \\
\midrule
\multicolumn{3}{c}{\emph{Signal construction parameters}} \\
\midrule
$n_{\text{diff}}$ & Derivative lookback window & $\{2\}$ \\
$w_{\text{ma}}$   & Derivative smoothing window & $\{2\}$ \\
$\lambda_1$ & Gating sensitivity (signal channel) &
$\{0.01,\;0.5,\;1,\;1.5\}$ \\
$\lambda_2$ & Gating sensitivity (derivative channel) &
$\{0.01,\;0.5,\;1,\;1.5\}$ \\
$A$ & Derivative amplitude scale &
$\{0.75,\;1,\;2\}$ \\
\midrule
\multicolumn{3}{c}{\emph{Walk-forward window parameters}} \\
\midrule
$w_{\text{fit}}$ & Training window length (minutes) &
{\{720, 1440, 2880, \newline 7200, 12000\}} 
\\
$\rho$ & Training--validation ratio &
$\{2,\;3,\;5,\;6\}$ \\
\bottomrule
\end{tabular}
\end{table}

These parameter ranges were selected based on preliminary sensitivity analyses
(Section~\ref{subsec:signal_scaling_gating}) to cover a wide but computationally
tractable search space.
The fixed values for \(n_{\text{diff}}\) and \(w_{\text{ma}}\) reflect the minimal
configuration that provides stable derivative estimates while preserving
high-frequency signal dynamics.


\subsection{Choice of the Hysteresis Threshold $\theta$}
\label{subsec:theta_selection}

The hysteresis threshold $\theta$ governs regime transitions in the final trading
The choice is closely related to the empirical
scale of the constructed signal.
Figure~\ref{fig:median_F_vs_parameters} characterizes how the typical magnitude of
the forward-oriented observable $F(t)$ varies as a function of the gating parameters,
summarized through $\mathrm{median}(|F(t)|)$.
This analysis identifies a natural signal scale of order unity across assets and parameter configurations.

Operating directly at the level of the median signal magnitude would, however, result in excessive regime switching.
Since $|F(t)|$ exceeds its median value by construction approximately half of the time, setting $\theta$ near the median would induce frequent threshold crossings and a high turnover trading regime dominated by short-lived fluctuations.
From a trading perspective, the hysteresis threshold must therefore be chosen above the typical median scale in order to enforce regime persistence and suppress noise-driven transitions.

Accordingly, we consider a discrete set of threshold values
\[
\theta \in \{0.6,\;0.8,\;1.0,\;1.4,\;1.6\},
\]
which spans values below, near, and above the median signal magnitude reported in Figure~\ref{fig:median_F_vs_parameters}.
This range allows us to study the sensitivity of trading behavior to the degree of hysteresis while remaining within a scale-consistent operating regime.

Importantly, the interpretation of Figure~\ref{fig:median_F_vs_parameters} does not
rely on forward-looking information.
While the figure is computed over a long sample for descriptive purposes, additional diagnostic checks (not shown) indicate that the empirical distribution of $|F_0(t)|$—and in particular its high percentiles—remains remarkably stable over
time for both assets considered.
This temporal stability implies that the characteristic scale of the signal is a
structural property of the normalization and aggregation procedure, rather than an
artifact of a specific market phase or sample endpoint.

\subsection{Equity Curves, Trading Activity, and Risk Characteristics}
\label{subsec:equity_and_risk}

\begin{figure*}[!ht] 
\centering
\includegraphics[width=\linewidth]{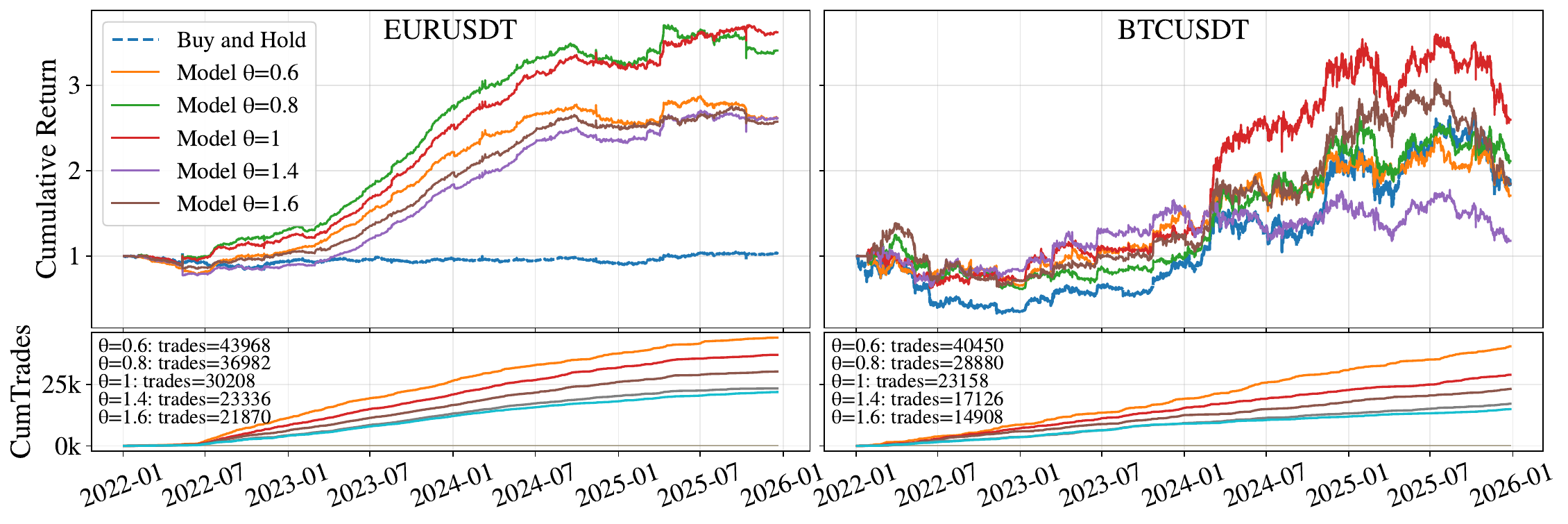}  

\caption{Walk-forward out-of-sample performance under zero transaction costs. Strategy versus Buy-and-Hold for different values of~$\theta$, and the cumulative number of executed trades.}
\label{fig:equity_and_trades}
\end{figure*}

Figure~\ref{fig:equity_and_trades} reports the out-of-sample walk-forward
performance of the proposed strategy for EURUSDT and BTCUSDT under zero
transaction costs.
For each asset, the figure compares the cumulative return of the proposed
strategy across different values of the threshold parameter~$\theta$ with a
Buy-and-Hold benchmark, and displays the cumulative number of executed trades
over time.

The equity curves reveal distinct but economically meaningful effects across
assets.
For EURUSDT, the proposed strategy exhibits sustained growth and consistently
outperforms Buy-and-Hold across a wide range of~$\theta$ values.
For BTCUSDT, although Buy-and-Hold attains a higher terminal return for some
thresholds, the proposed strategy produces markedly smoother equity curves,
with visibly reduced drawdowns and lower path variability.

The lower panels of Figure~\ref{fig:equity_and_trades} show that trading activity
differs across markets, reflecting differences in volatility and regime
dynamics rather than changes in signal construction, which is identical across
assets.
Importantly, variations in trading frequency across~$\theta$ do not translate
into increased instability of the equity process.
Even in the more volatile BTCUSDT market, the strategy maintains controlled
exposure and avoids the large drawdowns observed under Buy-and-Hold.

To quantify these qualitative observations, Table~\ref{tab:performance_metrics}
reports risk and return metrics for Buy-and-Hold and for the proposed strategy evaluated at the selected threshold value~$\theta^\ast = 1.0$.
This choice is not motivated by a maximization of out-of-sample performance.
Rather, $\theta^\ast= 1.0$ provides a convenient and interpretable reference point at which trading activity, regime persistence, and signal magnitude are all clearly expressed, facilitating the analysis of parameter distributions, trade statistics, and regime behavior.

Importantly, the qualitative conclusions drawn in the subsequent sections do not depend on this specific value.
As illustrated in Figure~\ref{fig:equity_and_trades}, alternative threshold choices lead to predictable changes in trading intensity but preserve the core behavior of the strategy.

\begin{table*}[!ht]
\centering
\footnotesize
\caption{Out-of-sample trading metrics for Buy \& Hold and the proposed strategy under the walk-forward framework.
Performance is computed by concatenating out-of-sample test windows across the full sample, with risk, drawdown, and activity metrics reported for $\theta^\ast = 1.0$.
All metrics are computed at the native one-minute frequency.
Sharpe and Sortino ratios are reported in non-annualized form.
For comparison with annualized benchmarks, these ratios can be scaled by
$\sqrt{365 \times 24 \times 60}$.
For standard definitions of the reported performance metrics, see, e.g.,~\citet{bacon2022risk}.}
\label{tab:performance_metrics}
\begin{tabular}{c l rr  @{\hspace{10mm}}  c l rr}
\toprule
\multicolumn{4}{c}{\textbf{EURUSDT}\ \  ($\theta^\ast = 1.0$)} & \multicolumn{4}{c}{\textbf{BTCUSDT}\ \  ($\theta^\ast = 1.0$)} \\
\cmidrule(lr){1-4}\cmidrule(lr){5-8}
Period & Metric & Buy \& Hold & Proposed & Period & Metric & Buy \& Hold & Proposed \\
\midrule
Jan 2022 & Total Return & 0.033304 & 2.620689 & Jan 2022 & Total Return & 0.849965 & 1.596073 \\
& Volatility & {\bf 0.000272} & {\bf 0.000181} & & Volatility & {\bf 0.000845} & {\bf 0.000649} \\
Dec 2025 & Downside Volatility & 0.000293 & 0.000266 & Dec 2025 & Downside Volatility & 0.000680 & 0.000688 \\
& Max Drawdown & -0.200671 & -0.114912 & & Max Drawdown & -0.677898 & -0.440523 \\
& Sharpe Ratio  & 0.000217 & 0.004859 & & Sharpe Ratio & 0.000908 & 0.001306 \\
& Sortino Ratio & 0.000201 & 0.003312 & & Sortino Ratio & 0.001128 & 0.001232 \\
& Calmar Ratio & 0.165963 & 22.805979 & & Calmar Ratio & 1.253825 & 3.623134 \\
& Ulcer Index & 9.630334 & 2.309119 & & Ulcer Index & 32.882349 & 16.633291 \\
& Time Under Water & 0.999107 & 0.966396 & & Time Under Water & 0.999351 & 0.983562 \\
& Total Trades & -- & 30{,}208 & & Total Trades & -- & 23{,}158 \\
& Trades per 1k candles & -- & \textbf{20.3} & & Trades per 1k candles & -- & \textbf{15.5} \\
\bottomrule
\end{tabular}
\end{table*}

Taken together, the results indicate that the proposed framework primarily acts
as a risk-reshaping mechanism.
By combining causal signal construction, adaptive derivative leading, and
hysteresis-based execution, the strategy systematically transforms raw market
exposure into a smoother equity process—sacrificing part of the upside in highly
volatile environments while delivering strong absolute and risk-adjusted
performance in more stable regimes.

\subsection{Trading Regime Duration and Position Holding Structure}
\label{subsec:regime_duration}

Beyond aggregate performance and risk metrics, the temporal structure of trading
regimes provides insight into the economic nature and practical viability of the
proposed strategy.
In particular, the distribution of position holding times reveals whether
performance is driven by sustained regime identification or by rapid oscillations
more likely associated with microstructure noise.

\begin{figure}[!ht]
\centering
\includegraphics[width=0.6\linewidth]{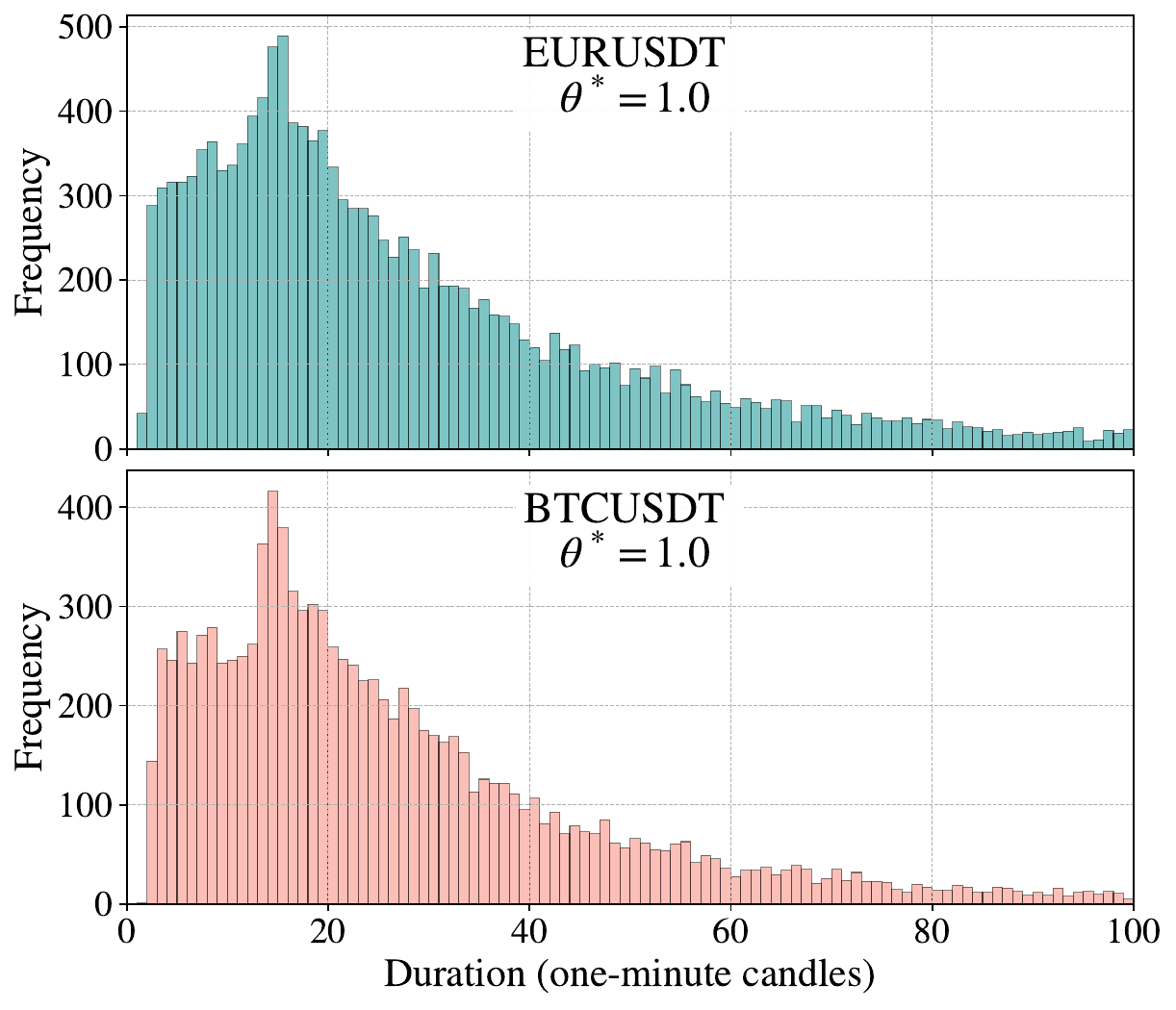}
\caption{Empirical distribution of position durations (non-flat segments) under the
walk-forward optimized strategy for EURUSDT and BTCUSDT. 
Both assets exhibit strongly right-skewed distributions, with a large mass of
short-lived positions and a non-negligible tail of persistent regimes.}
\label{fig:position_duration_hist}
\end{figure}

Figure~\ref{fig:position_duration_hist} reports the empirical distribution of
non-flat position durations (measured in one-minute candles) for EURUSDT and BTCUSDT
under the selected walk-forward configuration.
Each duration corresponds to a contiguous interval during which the strategy
maintains a long position before reverting to flat.

Both assets display highly asymmetric duration distributions, characterized by a
concentration of short holding times (on the order of tens of minutes) and a long
right tail extending to several thousand minutes.
This structure is consistent with frequent local regime fluctuations coexisting
with less frequent but persistent directional phases.

Despite substantial differences in volatility and market microstructure, the
qualitative shape of the duration distributions is remarkably similar across
EURUSDT and BTCUSDT.
This similarity suggests that the forward-oriented causal signal extracts regime
information in a manner that is robust across asset classes, rather than being
tuned to idiosyncratic price dynamics.
 
Table~\ref{tab:position_duration_stats} summarizes key descriptive statistics of the
position durations.
The median holding time is approximately 22--23 minutes for both assets, while the
upper quantiles reveal the presence of materially longer regimes.
In particular, the 90th percentile lies around 77 minutes, and the maximum durations
exceed 10{,}000 minutes in both markets, indicating that the strategy intermittently
locks into persistent regimes.

\begin{table}[!ht]
\centering
\footnotesize
\caption{Summary statistics of non-flat position durations (in one-minute candles) over the January 2022 to December 2025 period for $\theta^\ast = 1.0$.}
\setlength{\tabcolsep}{4pt}
\label{tab:position_duration_stats}
\begin{tabular}{ccccccc}
\toprule
Asset & Mean & Median & P25 & P75 & P90 & Max \\
\midrule
EURUSDT & 52.1 & 23 & 13 & 42 & 77 &  11{,}086 (05--17 Mar/2025)\\
BTCUSDT & 74.6 & 22 & 13 & 40 & 77 & 12{,}508 (17--30 Jan/2023)\\
\bottomrule
\end{tabular}
\end{table}

From a practical perspective, the prevalence of short-duration positions highlights the importance of transaction costs and execution frictions. Since all results are reported under zero transaction costs, the observed performance should be interpreted as a measure of the signal’s informational and regime-detection content, rather than as a directly tradable outcome.

At the same time, the presence of a non-negligible tail of long-duration positions indicates that the strategy is not merely exploiting high-frequency noise. Instead, it intermittently captures persistent regimes, precisely the environments in which transaction costs are less dominant and economic value is more likely to survive realistic execution conditions.

\subsection{Summary on Selected Hyperparameters Across Walk-Forward Epochs}
\label{subsec:selected_hyperparams}

\begin{figure}[!ht]
    \centering
    \begin{subfigure}[t]{0.6\linewidth}
        \centering
        \includegraphics[width=\linewidth]{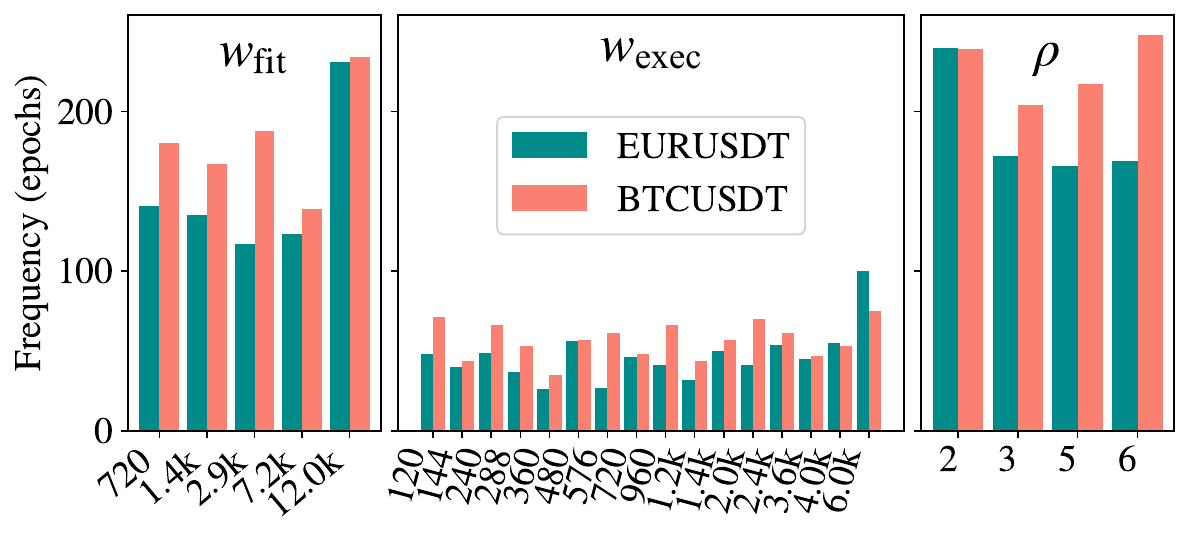} 
    \end{subfigure} 
    \vspace{-0.4em}  
    \begin{subfigure}[t]{\linewidth}
        \centering
        \includegraphics[width=0.6\linewidth]{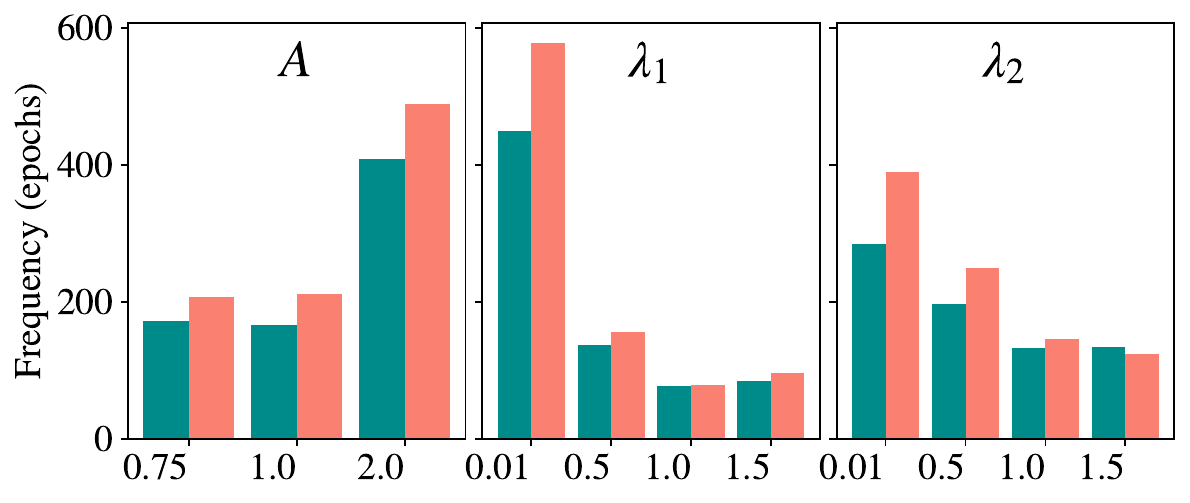} 
        \label{fig:selected_signal_params}
    \end{subfigure} 
    \caption{Empirical selection frequencies of hyperparameters under the walk-forward protocol.
    Each bar reports the number of epochs in which a given value is selected as optimal, shown
    separately for EURUSDT and BTCUSDT.}
    \label{fig:selected_hyperparams}
\end{figure}

To provide a compact view of the tuning behavior induced by the walk-forward selection procedure,
Figure~\ref{fig:selected_hyperparams} summarizes the empirical frequency with which each hyperparameter
value is selected as optimal across epochs, separately for assets.
Rather than reporting a single best configuration, these histograms reveal which regions of the search
space are repeatedly preferred by the out-of-sample validation criterion, hence offering an interpretable
diagnostic of hyperparameter stability under non-stationarity.

The top panels report the distribution of the rolling-window lengths used by
the walk-forward protocol: the training window length $w_{\mathrm{fit}}$, $w_{\mathrm{pred}}$ and the validation partition ratio
$\rho$. In our experiments, $w_{\mathrm{val}}=w_{\mathrm{pred}}$.
The bottom panel reports the distribution of the selected values for the
signal-construction parameters.

\section{Conclusion}
\label{sec:conclusion}
We introduced and empirically tested a strictly causal operator that transforms heterogeneous indicators into a forward-oriented decision observable in non-stationary markets.

The present framework is motivated by limitations observed in earlier fixed-parameter constructions, in which empirical performance was found to depend sensitively on market regime, particularly during extended low-volatility phases.
By introducing adaptive window calibration and scale-stable normalization, the approach proposed here is designed to mitigate this sensitivity by dynamically realigning signal dynamics with recent market conditions.

The observed behavior across assets reflects differences in the underlying market
structure. In EURUSDT, the strategy generates a stable cumulative return with
substantially reduced volatility and drawdowns relative to Buy-and-Hold. In BTCUSDT,
characterized by extreme volatility and deep benchmark drawdowns, the strategy
reshapes exposure by compressing downside risk and smoothing equity dynamics, at the
cost of reduced upside participation. 
These effects arise from systematic, causal adaptation to recent data rather than from any use of future information. At the same time, because parameters are selected from a finite grid based on past performance, some degree of local overfitting to the walk-forward objective remains possible.

At the same time, the effectiveness of the derivative-based leading mechanism relies
on implicit structural properties of the underlying signal. The anticipatory effect
is meaningful when the composite observable admits a locally smooth evolution, such
that a low-order expansion of the form
$F_0(t+\delta) \approx F_0(t) + \delta\,\partial_t F_0(t)$
provides a valid approximation. For signal geometries that deviate strongly from this
structure—e.g., highly irregular, discontinuous, or noise-dominated dynamics—the
derivative contribution may carry limited anticipatory information. Delineating the
class of signal behaviors for which causal derivative operators remain effective is an
important direction for future investigation.

A central limitation of the present study is the assumption of zero transaction costs.
Under this idealized setting, trading activity is high, and reported performance should
be interpreted as an upper bound. Nevertheless, the observed heavy-tailed distribution
of holding times, with a non-negligible population of long regimes, indicates that the
strategy intermittently captures persistent directional phases rather than reacting
exclusively to short-scale noise.

Future extensions should incorporate transaction costs and market impact directly into the walk-forward objective, potentially using regime duration as an explicit regularization variable. Systematic comparisons with high-capacity horizon-based models, including convolutional and recurrent architectures, would further clarify the trade-offs between causal signal engineering and deep predictive approaches. An additional priority is to replace the relatively broad hyperparameter grids used here by a more structured selection of a small number of parameter families, emphasizing robustness and interpretability over marginal in-sample performance gains; this is expected to reduce the effective degrees of freedom of the framework and hence its susceptibility to overfitting.

Overall, the results demonstrate that carefully designed causal operators can generate
effective forward-oriented structure under non-stationarity, offering an
interpretable and adaptive alternative to non-causal or purely data-driven trading
constructions.

Furthermore, all computations were implemented in Python using standard scientific libraries
(NumPy, pandas) and the code used for data processing and backtesting will be made available upon
request and released publicly after acceptance.



\section*{Declarations}

\subsection*{Funding}
The author received no external funding for this research.

\bibliography{references}

\begin{appendices}
\section{Technical Indicators and Composite Observable}
\label{app:indicators}
 
All indicators are computed using exclusively information available up to time~$t$, ensuring strict causality, and follow standard definitions widely used in empirical finance and technical analysis
(see~\citet{appel1979} and \citet{murphy1999technical}
for details).
The indicators  $\{Z_t^{(k)}\}$ are summarized in Table~\ref{tab:indicators}, together with their economic interpretation and defining equations. 
Each indicator captures a complementary aspect of market dynamics, as briefly summarized below.

\begin{table}[!ht]
\centering
\footnotesize 
\caption{Technical indicators used to build the composite observable $F_0(t)$.
RSI, MFI and BB\% are bounded in $[0,100]$, whereas MACD is scale-dependent.}
\label{tab:indicators}
\begin{tabular}{l c l}
\toprule
Indicator & Range & Definition \\
\midrule
RSI & $[0,100]$ &
$\displaystyle 100\times\!\left(1-\frac{1}{1+G_t/L_t}\right)$ \vspace{4pt}\\
MFI & $[0,100]$ &
$\displaystyle 100\times\!\left(1-\frac{1}{1+\mathrm{PMF}_t/\mathrm{NMF}_t}\right)$ \vspace{4pt}\\
MACD diff.\tnote{a} & scale-dependent &
$\displaystyle (\mathrm{EMA}_{\mathrm{fast}}-\mathrm{EMA}_{\mathrm{slow}})_t-\mathrm{Signal}_t$ \vspace{4pt}\\
BB\% & $[0,100]$ &
$\displaystyle 100\times\!\left(\frac{P_t-(\mu_t-k\sigma_t)}{(\mu_t+k\sigma_t)-(\mu_t-k\sigma_t)}\right)$ \\
\bottomrule
\end{tabular} 
\end{table}

The Relative Strength Index (RSI), following \citet{wilder1978}, measures the balance between recent gains and losses and provides a bounded representation of short-term momentum.
The Money Flow Index (MFI) extends this concept by incorporating traded volume, thereby capturing buying and selling pressure beyond price movements alone \cite{murphy1999technical}.
The MACD difference reflects changes in trend intensity by comparing short- and long-horizon exponential moving averages and their signal line \cite{appel1979}.
Finally, the Bollinger Band Percent (BB\%) expresses the relative position of the price within a volatility envelope, yielding a dimensionless measure of price extremeness conditional on recent volatility.

\section{Derivative-Based Signal Leading: Illustrative Examples}
\label{app:derivative_leading}

The forward-oriented operator introduced in Section~\ref{sec:F_def} augments the composite
signal with a causal derivative component in order to promote systematic phase advance
while preserving strict causality.
To illustrate the underlying mechanism in a controlled setting, we consider the
demonstrative signal
\begin{equation}
f(t) = \sin\!\big(a t + A_0 \sin(w t)\big) + m t - 1,
\label{eq:demonstrative_function}
\end{equation}
with $a=0.5$, $A_0=2$, $w=1$, and $m=0.1$.
This construction combines oscillatory behavior with slow modulation and a linear trend, reproducing qualitative features typical of non-stationary signals.
The causal derivative is approximated via backward finite differences,
\begin{equation}
\frac{df}{dt} \approx \frac{f(t) - f(t-n\Delta t)}{n\Delta t},
\label{eq:causal_derivative}
\end{equation}
where $\Delta t$ denotes the sampling interval and $n=5$ for this example.

\paragraph{Derivative leading and amplitude control}

We first isolate the basic leading effect obtained by combining the signal with its
scaled causal derivative,
\begin{equation}
f_{\text{lead}}(t) = f(t) + A \cdot \frac{df}{dt}.
\label{eq:leading_composite}
\end{equation}

\begin{figure}[!ht]
\centering
\includegraphics[width=0.6\linewidth]{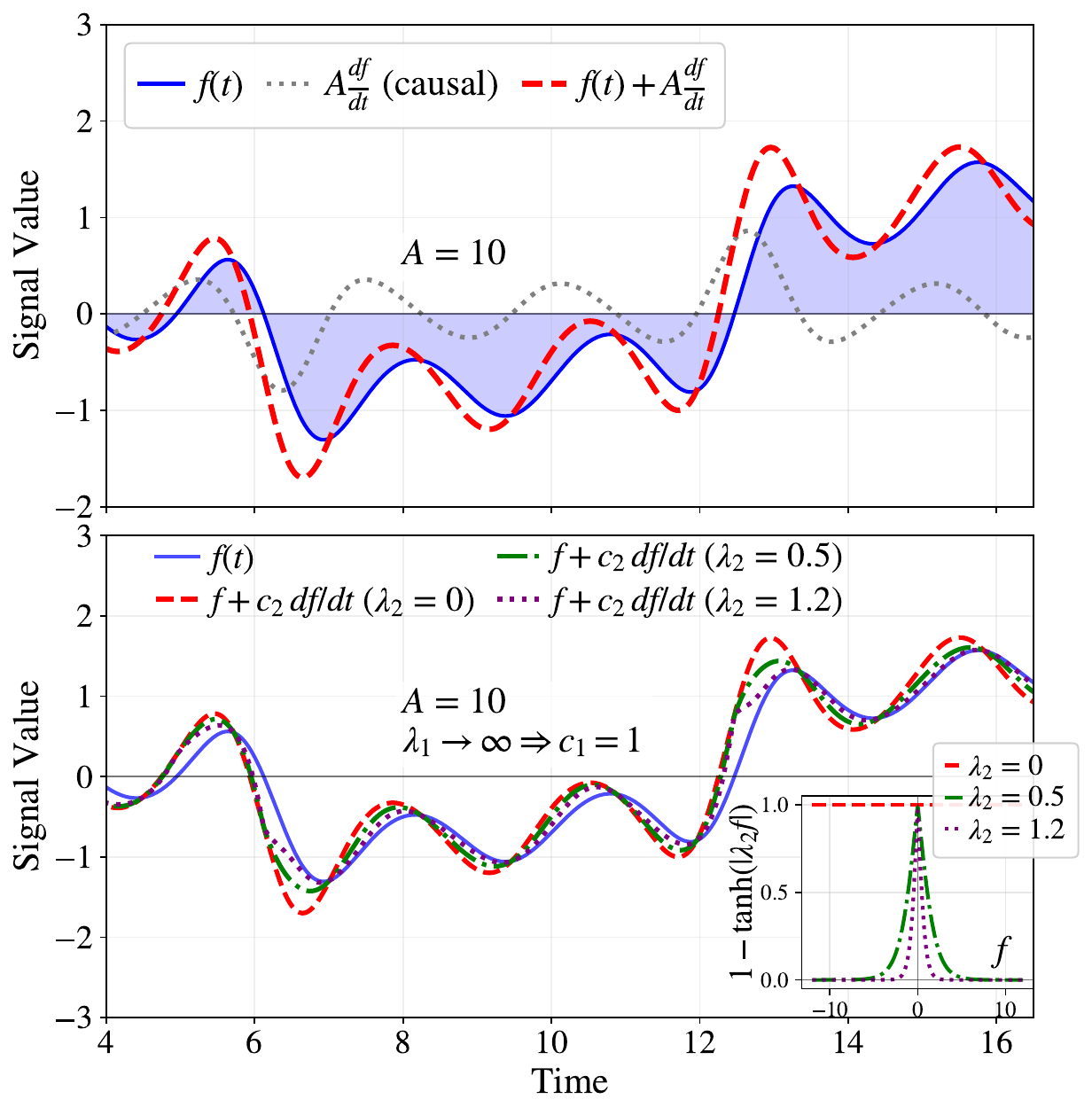} 
\caption{Derivative-based signal leading under strict causality.
\textbf{Top:} The signal $f(t)$ combined with its scaled causal derivative,
$f(t)+A\,df/dt$, exhibits a systematic phase advance, most visible near zero-crossings.
\textbf{Bottom:} Adaptive gating of the derivative term for different $\lambda_2$
with $A=10$ and $\lambda_1 \to \infty$ ($c_1=1$). Increasing $\lambda_2$ suppresses
the derivative contribution away from zero-crossings. For $\lambda_2=0$, the gated
signal coincides with the derivative-enhanced signal shown in the top panel.} 
\label{fig:derivative_and_gating}
\end{figure}

The top panel of Figure~\ref{fig:derivative_and_gating} shows that the derivative term
induces a systematic phase advance relative to $f(t)$.
This effect is most pronounced near zero-crossings, where the local slope is largest.
The amplitude parameter $A$ controls the strength of the leading behavior: increasing $A$
yields stronger anticipation but also amplifies high-frequency fluctuations. 

\paragraph{Adaptive gating via $\lambda$ parameters.}

In financial time series, derivative information is not uniformly informative across
regimes.
This motivates the adaptive gating mechanism,
\begin{equation}
f_{\text{out}}(t) = c_1(t)\,f(t) + c_2(t)\, {f'(t)},
\end{equation} 
with,
\begin{equation}
c_1(t) = \tanh\left(|\lambda_1 f(t)|\right); \ \ \ 
c_2(t) = A\big[1 - \tanh\left(|\lambda_2 f(t)|\right)\big].
\end{equation}

The bottom panel of Figure~\ref{fig:derivative_and_gating} illustrates the role of the
parameter $\lambda_2$ for fixed $A=10$.
As $\lambda_2$ increases, the derivative contribution is progressively suppressed when
$|f(t)|$ is large (strong trends) and concentrated near zero-crossings (regime
transitions).
This mechanism focuses the leading effect precisely where anticipation is most valuable,
while avoiding excessive amplification during persistent trends.
Importantly, the figure also clarifies the role of $\lambda_1$.
In the limit $\lambda_1 \to \infty$, the coefficient $c_1(t)$ converges to unity for all
$t$, and the gated construction reduces to
\[
f_{\text{out}}(t) = f(t) + c_2(t)\, {f'(t)}.
\]
In this regime, the red curve in the right panel corresponding to $\lambda_2=0$
coincides exactly with the red curve in the left panel, which represents the ungated
derivative-enhanced signal $f(t) + A\,df/dt$.
Thus, the left panel can be interpreted as a limiting case of the full gated model,
providing a clean reference for understanding how adaptive gating modulates the leading
effect.

These illustrative examples motivate the structure of the empirical trading signal used
in this study:
(i) causal derivatives can induce systematic phase advance without look-ahead bias;
(ii) the amplitude parameter $A$ controls the strength of this anticipation;
(iii) the gating parameters $\lambda_1$ and $\lambda_2$ regulate when derivative
information is incorporated, emphasizing regime transitions while suppressing noise
during strong trends.

\end{appendices}
\end{document}